\documentclass[10pt]{article} 
\usepackage{amssymb,epsf}
\usepackage{amsmath}
\usepackage[dvips]{graphicx}
\usepackage[dvips]{color}
\usepackage[a4paper]{geometry}
\usepackage{subfigure}
\usepackage{setspace}

\newcommand\BE[1]{{\begin{equation}#1\end{equation}}}

\newcommand\PHI[2]{\Phi_{#1}{#2}}

\newcommand{\DD}{{\cal D}}

\newcommand\NN[1]{{\cal N}_{#1}}
\newcommand\MM[1]{{\cal M}_{#1}}

\newcommand\union\cup

\newcommand\EPS{{\mu_0 { Y}^2}}
\newcommand\PP{{}^\prime}

\newcommand{\EM}{ electromagnetic }
\newcommand{\beq}{\begin{equation}}
\newcommand{\beqa}{\begin{eqnarray}}
\newcommand{\eeq}{\end{equation}}
\newcommand{\eeqa}{\end{eqnarray}}
\newcommand{\non}{\nonumber}
\newcommand{\fr}[1]{(\ref{#1})}

\def\PP{{\cal P}}

\def\UU{{\cal U}}

\def\WW{{\cal W}}

 \def\ee{{\mathbf e}}

 \def\hh{{\mathbf h}}

\def\hhbar{{\bar{\hat{\mathbf h}}}}

\def\dUU{{\frac{d\UU}{d\omega}}}

\newcommand\PSII[2]{{\Psi}_{#1}{#2}}
\newcommand\PHII[2]{{\Phi}_{#1}{#2}}

\newcommand\PSIB[2]{\overline{\Psi_{#1}}{#2}}
\newcommand\PHIB[2]{\overline{\Phi_{#1}}{#2}}

\def\AA{{\kappa}}
\def\LL{{\cal L}}
\def\NN{{\cal N}}
\def\EE{{\cal E}}
\def\OO{{\cal O}}
\def\EEE{E}
\def\er{\hat{{\mathbf e}}_r}
\def\ez{ \hat{{\mathbf e}}_z }
\def\eth{ \hat{{\mathbf e}}_\theta }

\def\hatlap{\hat {\nabla^2}}

\begin{document}

\title{ Energy Spectra from Electromagnetic Fields Generated by Ultra-relativistic Charged Bunches in a Perfectly Conducting Cylindrical Beam Pipe}

\author{Alison C Hale\,\footnote{a.c.hale@lancaster.ac.uk}
and
Robin W Tucker\,\footnote{r.tucker@lancaster.ac.uk}\\
Department of Physics, Lancaster University\\
and the Cockcroft Institute,\\ Keckwick Lane,\\ Daresbury,
WA4 4AD, UK
}

\date{\today}


\maketitle
\begin{abstract}
The spectrum of electromagnetic fields    satisfying
perfectly conducting boundary conditions in a segment of a straight beam pipe with a circular
cross-section is discussed as a function of various  source models.  These include charged bunches that  move along the axis of the pipe with constant speed
for which an exact solution to the initial-boundary value problem for Maxwell's equations in the beam
pipe is derived. In the ultra-relativistic limit all longitudinal components of the fields tend to
zero and the spectral content of the transverse fields and average total electromagnetic energy
crossing any section of the beam pipe are directly related to the properties of the
ultra-relativistic source. It is shown  that for axially
symmetric ultra-relativistic bunches interference effects occur that show a striking resemblance
to those that occur due to CSR in cyclic machines despite the fact that in this limit the source
is no longer accelerating. The results offer an analytic description showing  how such enhanced
spectral behaviour depends on the geometry of the source and the details of  the stochastic
distribution of structure within the source. The field energy spectra associated  with a source  containing  ${\cal N}$ identically  charged ultra-relativistic pulses, each with individual longitudinal gaussian profiles  distributed according to  a uniform probability distribution with compact support, is compared with that  generated by charged bunches containing a  distribution with $2n+1$ peaks  in  a region with compact support (modelling micro-bunches).  These results
offer a viable experimental means for inferring properties of the longitudinal charge distribution
of bunches with micro-structure in  ultra-relativistic motion in straight segments of a beam pipe
from observation of the associated electromagnetic energy spectra. They are also of relevance to
design criteria where the coherence effects of such fields play a significant role.
\bigskip
\end{abstract}
%
%
%
%
%
%
%
\maketitle
\section{Introduction}

Modeling the behaviour of charged particles in a modern accelerator
is a critical component in its design. Due to the inherent
non-linear nature of the dynamics of charged bunches in external
electromagnetic fields such modeling generally necessitates numerical
computation. With current hardware such computations often require
further stringent approximations on the equations of motion in order
to extract viable information. To date the most sophisticated
Maxwell-Vlasov solvers are unable to effectively  model
electromagnetic interference in 3 spatial dimensions.
Approximations often neglect the effects of confining boundaries and
wakes,  radiation reaction forces,  detailed stochastic properties
of the beam and possible quantum effects.  However, in certain
limits {\it analytic} information can  be deduced from the
fundamental equations of motion for the coupled particle field
system.  If the motion is prescribed the problem reduces to solving
Maxwell's equations for convected sources. Nevertheless finding solutions
satisfying  boundary condition appropriate to an accelerator is in
general a non-trivial exercise.

 Since Fourier analysis is a linear operation the spectral content of the electromagnetic energy in the fields produced by charged particles in prescribed motion {\it in free space}  has been exhaustively investigated over many decades, particularly for collections of particles in uniform circular motion \cite{schwinger_1949,schwinger_1946}.  The effects of the superposition of retarded free-space solutions to Maxwell's equations with distributed currents on the spectral content of their radiation fields gave rise to the notion of {\it coherent synchrotron radiation} (CSR) in which enhanced radiation in some frequency domain can occur,  depending quadratically on the total number of point particles in an accelerating bunch \cite{hofmann}.  It is generally assumed that the effects of confining boundaries in a real cyclic machine do not significantly alter the {\it criteria for CSR} from those that arise for sources in free space. However analytic efforts to determine such criteria  taking account of boundary conditions and stochastic effects inevitably demand further approximations \cite{saxon_1954,saldin}. Since the role of CSR is a fundamental ingredient in the design of new light sources it is of interest to explore new analytic approximations schemes that can complement the numerical simulations of particle-field interactions.

In this note we approach the modeling process by ensuring at the outset that all fields satisfy
perfectly conducting boundary conditions in a segment of a straight beam pipe with a circular
cross-section.  The sources will be assumed to move along the axis of the pipe with constant speed
and an exact solution to the initial-boundary value problem for Maxwell's equations in the beam
pipe is derived. In the ultra-relativistic limit all longitudinal components of the fields tend to
zero and the spectral content of the transverse fields and average total electromagnetic energy
crossing any section of the beam pipe are directly related to the properties of the
ultra-relativistic source. Thus a measurement of the former offers a direct method of estimating
properties of the latter see e.g. \cite{jamison}. However the detailed structure of the spectral
content depends on the source structure. In the following we demonstrate that for axially
symmetric ultra-relativistic bunches interference effects occur that show a striking resemblance
to those that occur due to CSR in cyclic machines despite the fact that in this limit the source
is no longer accelerating. Furthermore it is possible to study analytically how such enhanced
spectral behaviour depends on the geometry of the source and the details of  the stochastic
distribution of structure within the source.

Section 2 establishes the formalism  for solving Maxwell's equations for fields inside  a perfectly conducting cylindrical beam pipe in terms of  complex Dirichlet and Neumann eigen-modes of the 2-dimensional scalar Laplacian operator. An exact solution to the initial-boundary value problem  is exhibited and this is used to explore fields associated with ultra-relativistic bunches.
Section 3 develops the model in terms of an axially symmetric source, offers a compact formula for the spectral distribution of electromagnetic energy that crosses any section of the beam pipe and, in section 4,  discusses the effects on the spectra produced by different types of stochastic distribution.

\section{{Electromagnetic Field Solutions}}

The electromagnetic fields ${\mathbf e} $ and $
{\mathbf h} $  in a beam pipe in the presence of sources
 with charge density $\rho$ and electric current density ${\mathbf J}$
satisfy the Maxwell system:

$$ \nabla \times {\mathbf e } + \mu_0\, \partial_t{\mathbf h} =0 $$

$$\nabla \times \,{\mathbf h} - \,\mu_0 Y^2 \,\partial_t{\mathbf e} -{\mathbf  J}=0$$

$$ \nabla\cdot \,{\mathbf h}=0$$

$$\mu_0 Y^2\, \nabla\cdot \,{\mathbf e} - \rho=0$$

where the admittance $Y= 1/(\mu_0 c)$ with $c$ being the speed of light in vacuo. All vectors will be referred to a global ortho-normal frame $\{ \er,\ez,\eth  \}$ defining a cylindrical coordinate system   $\{r,\theta,z\}$ with the axis of the beam pipe along the $z-$axis.

A transverse circular section of this pipe is denoted  ${\cal D}$ with boundary $\partial{\cal D}$. In the following we exploit the properties of a
  {\it complex Dirichlet mode set} $\{ \Phi_N\}$. This  is a
collection of complex eigen-functions of the 2-dimensional (transverse)  Laplacian operator
$\hatlap  $ on ${\cal D}$
that vanishes on $\partial{\cal D} $.
This boundary condition determines the
associated (positive non-zero real) eigenvalues $\beta_N^2$. The
label $N$ here consists of an ordered pair of real numbers. Thus

\BE{ \hatlap \PHII N {} - \beta_N^2 \PHII N {} =0, \label{dmode}} with $\PHII N {}
\vert_{\partial{\cal D}}=0.$ With an overbar indicating complex conjugation these modes are
normalised to satisfy \beq \int_{{\cal D}}\PHIB
 M {}\, \PHI N {} \,r\,dr\,d\theta=
{\cal N}_N^2\,\delta_{NM},
\label{eqn:Dirichelet-orthogonal}
\eeq

An explicit form for
$\Phi_N$ is for $n\in\mathbb{Z}$
\beq
\Phi_N(r,\theta)=J_n\left(x_{q(n)}\frac{r}{a}\right) e^{in\theta},
\label{eqn:Phi-basic}
\eeq
where $J_n(x)$ is the $n$-th order Bessel function
and the numbers $\{x_{q(n)}\}$ are defined by $J_n(x_{q(n)})=0$ and
$N:=\{n,q(n)\}$. The eigenvalues are given by
$\{\beta_N=x_{q(n)}/a\}$ and
 $\NN^2_N=\pi a^2 J^2_{n+1}(x_{q(n)})$.

In a similar manner   a {\it Neumann mode set} $\{\Psi_N\}$ is a collection of eigen-functions of
the Laplacian operator  $\hatlap  $ on ${\cal D}$ such that $\frac{\partial \Psi_N } {\partial r}$
vanishes on $\partial{\cal D} $. This alternative boundary condition determines the associated
(positive non-zero real) eigenvalues $\alpha_N^2$ where again the label $N$ consists of an ordered
pair of real numbers: \beq \hatlap\PSII N {} - \alpha_N^2 \PSII N {}=0, \label{nmode} \eeq with $
\frac{ \partial\Psi_N }{\partial r} \vert_{\partial{\cal D}}=0$. These modes are normalized to
satisfy
$$
\int_{{\cal D}}\PSIB M {}\, \Psi_N {} \,r\,dr\,d\theta={\cal M}_N^2\,\delta_{NM}.
$$

An explicit
form for  $\Psi_M$ is  for $m\in\mathbb{Z}$
\beq
\Psi_M(r,\theta)=J_m\left(x'_{p(m)}\frac{r}{a}\right) e^{im\theta},
\label{eqn:Psi-basic}
\eeq
where the numbers  $\{x'_{p(m)}\}$ are defined by $J_m'(x'_{p(m)})=0
$ and $M:=\{m,p(m)  \}$. The eigenvalues are given by
$\{\alpha_M=x_{p(m)}'/a\}$ and
 $\MM{M}^2=\pi a^2 J^2_{m+1}(x'_{p(m)})$.

The electromagnetic fields in the interior of the cylindrical beam pipe of radius $a$ satisfying
perfectly conducting boundary conditions at $r=a$ can now  be expanded
\cite{jones,tucker1,tucker2} as:

 \BE{\ee(t,z,r,\theta)= \sum_N V^E_N(t,z)\, \nabla \Phi_N(r,\theta) + \sum_M V^H_M(t,z) \,  \hat{\mathbf e}_z\times \nabla \Psi_M(r,\theta)  + \sum_N \gamma^E_N(t,z)\, \Phi_M(r,\theta) \,\hat{\mathbf e}_z \label{expande}}

\BE{\hh(t,z,r,\theta)= \sum_N I^E_N(t,z)\, \hat{\mathbf e}_z\times \nabla \Phi_N(r,\theta) + \sum_M I^H_M(t,z) \,  \nabla\Psi_M(r,\theta) + \sum_M \gamma^H_M(t,z)\, \Psi_M(r,\theta) \,\hat{\mathbf e}_z\label{expandh}}

The fields are assumed to be generated by an external RF  source that accelerates charged bunches to near the speed of light. In a straight beam pipe  such a source can be  modeled  by an arbitrary smooth {\it convective} charge density $\rho$ and a current with ortho-normal components
 $J_r=J_\theta=0,\, J_z(z-vt,r,\theta)=v\,\rho(z-vt,r,\theta)$ with $v$ close to the speed of light.
 The equations for
$\gamma_N^{H}$ and $\gamma_N^{E}$  that follow from Maxwell's equations  and (\ref{expande}),
(\ref{expandh})  for these sources  are:
 \beq
\ddot{\gamma}_N^{H}-c^2\gamma_N^{H\prime\prime} +c^2\alpha_N^2\gamma_N^{H}=0
\label{eqn:Sol-gammaH:0} \eeq \beq \ddot{\gamma}_N^{E}-c^2\gamma_N^{E\prime\prime}
+c^2\beta_N^2\gamma_N^{E}=-\frac{c^2\mu_0 }{\NN^2_N}(c^2 - v^2) \overline{\rho^{\prime}_N}
\label{eqn:Sol-gammaE:0} \eeq

where

\beq
\rho_N:=\int_\DD\rho\,\Phi_N\,r\,dr\,d\theta.
\label{eqn:def-rho-j-N}
\eeq

In terms of $\gamma_N^{H}$ and $\gamma_N^{E}$ and the
projected convective  sources
\beqa V_N^{E}&=&\frac{1}{\beta_N^2}\bigg(
\gamma_N^{E\prime}-\frac{1}{\NN^2_N\EPS}\overline{\rho^{}_N}
\bigg),
\label{eqn:Sol-VE:0}\\
V_N^{H}&=&\frac{\mu_0 }{\alpha_N^2}\dot{\gamma}_N^{H},
\label{eqn:Sol-VH:0}\\
I_N^{E}&=&-\frac{1}{\beta_N^2}\bigg( \EPS\dot{\gamma}_N^{E}
+\frac{v}{\NN^2_N} \overline{\rho^{}_N}\bigg),
\label{eqn:Sol-IE:0}\\
I_N^{H}&=&\frac{1}{\alpha_N^2}\gamma_N^{H\prime}. \label{eqn:Sol-IH:0} \eeqa

Finding the fields $\ee$ and $\hh$ is now reduced
 to solving an initial-value problem for the
decoupled fields
$\gamma_N^{H},\gamma_N^{E}$. For some real constant $\sigma>0$  and source $g(t,z)$  each is a solution to  the generic
 (hyperbolic) partial differential equation
\beq
\ddot{f}-c^2
f^{\prime\prime}+c^2\sigma^2 f=g.
\label{eqn:typical-diff-in-z}
\eeq
The general causal solution $f(t,z)$ with prescribed values of $f(0,z)$ and
$\dot{f}(0,z)$, is (see e.g. \cite{pinsky})
\beq
f(t,z)={\cal H}_{\sigma}[f^{init}](t,z)
+{\cal I}_{\sigma}[g](t,z),
\label{eqn:Pinsky}
\eeq
where
\beqa
&&{\cal H}_{\sigma}[f^{init}](t,z):=
\frac{1}{2}\bigg\{f(0,z-ct)+f(0,z+ct)\bigg\}\non\\
&& +\frac{1}{2c}\int_{z-ct}^{z+ct}d\zeta\,
\dot{f}(0,\zeta)J_0(\sigma\sqrt{c^2t^2-(z-\zeta)^2})\non\\
&& -\frac{ct\sigma}{2}\int_{z-ct}^{z+ct}d\zeta\, f(0,\zeta)
\frac{J_1(\sigma\sqrt{c^2t^2-(z-\zeta)^2}\,)}{\sqrt{c^2t^2-(z-\zeta)^2}},
\label{eqn:def-H}
\eeqa
and
\beq
{\cal I}_{\sigma}[g](t,z):=\frac{1}{2c}\int_0^t
dt'\int_{z-c(t-t')}^{z+c(t-t')}d\zeta\,
g(t',\zeta)J_0(\sigma\sqrt{c^2(t-t')^2-(z-\zeta)^2}),
\label{eqn:def-I}
\eeq
The functions $f(0,z), \dot{f}(0,z)$
constitute the initial $t=0$ Cauchy data in this solution and
determine the ${\cal H}_\sigma$ contribution above.

For a bunch with total charge $Q$ moving with speed $v$ we assume here that $\rho$ can be written
$$
\rho(z-vt,r,\theta)=Q\,
\rho^{\perp}(r,\theta)\rho^{\parallel}(z-vt),
$$
where $v$ is given ($v\leq c$) and
$\rho^{\perp}(r,\theta)$,  $\rho^{\parallel}(z-vt)$ are
arbitrary smooth functions subject to
$$
\int_\DD \rho^{\perp}(r,\theta)\,r\,dr\, d\theta=1,\qquad
\int_{-\infty}^{\infty}dz\, \rho^{\parallel}(z-vt)=1,
$$
With these sources the causal solutions to \fr{eqn:Sol-gammaH:0} and \fr{eqn:Sol-gammaE:0} for $\gamma_N^{H}$ and $\gamma_N^{E}$
are given by:
\beqa
\gamma_N^{H}(t,z)&=&{\cal H}_{\alpha_N}[\gamma_N^{H init}] (t,z),
\non\\
\gamma_N^{E}(t,z)&=&{\cal H}_{\beta_N}[\gamma_N^{E init}] (t,z)
-\frac{\mu_0  c^2}{\NN^2_N}(c^2-v^2){\cal I}_{\beta_N}
[\overline{\rho_N^{\prime}}] (t,z),\non
\eeqa
where  $\rho_N^{}$ is given by
\fr{eqn:def-rho-j-N} and
\beq
{\rho_N^{\,\prime}}(t,z)=Q\,\rho^{\parallel\prime}(z-vt)
\int_\DD\rho^{\perp}(r,\theta)
{\Phi_N}\, r\,dr\,d\theta.
\label{eqn:rho-bar-N-0-loc}
\eeq
In the ultra-relativistic limit, $v\rightarrow c$,
the second term in
$\gamma_N^{E}(t,z)$ tends to zero.

\section{{Spectral Energy Distributions}}

When the transverse
distribution depends only on $r$,
expressions for the electromagnetic fields  associated with the sources simplify.
The source under consideration is axially symmetric if
\beq
\rho^{\perp}(r,\theta)={\cal R}(r),
\label{eqn:axial-dist}
\eeq
where ${\cal R}(r)$ is a smooth function satisfying
\beq
\int_0^{a}dr\, r{\cal R}(r)=\frac{1}{2\pi}.
\label{eqn:axial-dist-norm}
\eeq
Then for axially symmetric bunches
\beq
{\rho_N^{\,\prime}}(t,z)=2\pi Q_{tot}\delta_{n,0}
\rho^{\parallel\prime}(z-vt)
\int_0^a dr\, r{\cal R}(r)J_0\left(x_{q(0)}\frac{r}{a}\right).
\label{eqn:symmetric-dist-source}
\eeq

From these formulae  the instantaneous {\it real} \EM power   crossing any section of the beam
pipe is obtained by integrating the component  ${S}= (\ee \times \hh) \cdot {{\mathbf e_z}} $ of
the Poynting vector field   over the cross-section $\DD$ at an arbitrary point with coordinate $z$
$$
\PP(t,z):=\int_\DD { S }(t,z,r,\theta)\,r\,dr\,d\theta
$$

With the aid of the orthogonality properties of the Dirichlet and Neumann modes this becomes
\beq
\PP=\Re\bigg\{\sum_N \beta_N^2\NN^2_N
V_N^{E}\overline{I_N^{E}} -\sum_M \alpha_M^2\MM{M}^2
V_M^{H}\overline{I_M^{H}} \bigg\},
\label{eqn:pow-0th-result} \eeq
where $\Re$ takes the real part of
its argument. From
\fr{eqn:pow-0th-result}, \fr{eqn:Sol-VE:0},
 \fr{eqn:Sol-VH:0}, \fr{eqn:Sol-IE:0}, and
\fr{eqn:Sol-IH:0}  this  power flux can be explicitly expressed
in terms of the source projections $\rho_N$, ${\gamma}_M^{H}$ and ${\gamma}_N^{E}$
\beqa
\PP&=&\Re\bigg\{\sum_N\bigg( \frac{v
|\rho^{}_N|^2}{\NN^2_N\beta_N^2\EPS}
+\frac{\overline{\dot{\gamma}_N^{E}}}{\beta_N^2}
\overline{\rho^{}_N} -\frac{v\gamma_N^{E\prime}}{\beta_N^2}
\rho^{}_N\non\\
&&-\frac{\EPS}{\beta_N^2}\NN^2_N
\gamma_N^{E\prime}\overline{\dot{\gamma}_N^{E}} \bigg)
-\mu_0 \sum_M\frac{\MM{M}^2}{\alpha_M^2}\dot{\gamma}_M^{H}
\overline{\gamma_M^{H\prime}}\bigg\}.
\label{eqn:pow-0th-tra-charge}
\eeqa
In  a cyclic machine with time-periodic fields of period $T$ (such as a synchrotron) the fields associated with the sources can be expanded in a Fourier series:
$$ \hat{\ee_n}(z,r,\theta) = \frac{1}{T} \int _0^T \ee(t,z,r,\theta) \exp( i\Omega t  )\,dt $$
$$ \hat{\hh_n}(z,r,\theta) = \frac{1}{T} \int _0^T \hh(t,z,r,\theta) \exp( i\Omega t  )\,dt $$
 where $\Omega=2\pi/T$,
and the mean (time-averaged) power crossing $\DD$  at $z$ is
$$ < \PP>(z) = \sum^\infty_{n=-\infty} <\PP_n>(z)$$ where

$$ <\PP_n >(z)= \int_\DD  (\hat\ee_n(z,r,\theta) \times  {\hhbar_n} (z,r,\theta)) \cdot  {\hat{\mathbf e_z}}) r\,dr\,d\theta$$
However in  a straight beam pipe the fields associated with the sources are not periodic in time. The sources are however localized in space so the  total electromagnetic energy  $\UU$ crossing any section $\DD$ at $z$ is well-defined:
$$\UU(z)= \int^\infty_{-\infty} \PP(t,z)\,dt= \int^\infty_{-\infty} \left(  \frac{d\UU}{d\omega}(\omega,z) \right) \,d\omega$$
where
 \BE{\frac{d\UU}{d\omega}(\omega,z) = \int_\DD  \left(\hat{\ee}(\omega,z,r,\theta) \times  {\hhbar} (\omega,z,r,\theta) \cdot  {{\mathbf e_z}}\right) \,r\,dr\,d\theta \label{master}  }
in terms of the Fourier transforms:
$$ \hat\ee (\omega, z,r,\theta)= \frac{1}{\sqrt{2\pi}} \int^\infty_{-\infty} \ee(t,z,r,\theta) \exp(i\omega t)\,dt $$

$$ \hat\hh (\omega, z,r,\theta)= \frac{1}{\sqrt{2\pi}} \int^\infty_{-\infty} \hh(t,z,r,\theta) \exp(i\omega t)\,dt $$
Clearly $\int_{\omega_1}^{\omega_2} \frac{d\UU}{d\omega}(\omega,z)\,d\omega$ is the total electromagnetic energy crossing $\DD$   at the station $z$ of the beam pipe in the wave band $\omega_1$ to $\omega_2$ and is experimentally accessible  with suitable detector diagnostics.
 Measurement of  the spectral content of such energy associated with ultra-relativistic charged bunches can offer valuable information about their longitudinal charge distribution.

In  the ultra-relativistic  limit the fields $\ee$ and $\hh$ have
components that lie solely in the transverse  sections of the pipe
and are concentrated  in space where the sources are concentrated.
For an   axially symmetric  ultra-relativistic bunch with charge $Q$
one calculates from (\ref{master})
$$ \dUU = K \vert \hat{\rho}^{\parallel} (\omega,z) \vert^2$$
where the constant $$K= \frac{c}{\epsilon_0} Q^2 \sum_M \frac{\vert  \int_\DD {\cal R}(r)\Phi_M(r,\theta)\,r\,dr\,d\theta \vert^2}{\vert 2\pi \beta_M\int_0^a J_0^2(\beta_M\,r)\,r\,dr  )  \vert^2}$$

In general one expects the source to require a stochastic description since its origin is fundamentally stochastic. This requires the introduction of stochastic variables and their associated probability distribution in order to calculate the expectation values of $\dUU$. However it is of interest first to note the deterministic structure of $\dUU$ that arises from a well-defined finite train of $\NN$  charged ultra-relativistic  pulses where  each pulse has the same longitudinal profile  $f$ in $\rho^{\parallel}$. Denote the  longitudinal charge distribution  $\rho^{\parallel}$ of such a train by $F$ with
$$F(z-ct)= F_0\sum_{j=1}^\NN f(z-c(t-T_j))$$
for some real constant $F_0$.

If  $\hat{f}(\omega)$ is the Fourier transform of $f(t)$ then the Fourier transform of $\sum_{j=1}^\NN\,f(t+\tau_j)$ with respect to $t$ is $\hat{f}(\omega)\sum_{j=1}^\NN\,\exp(-i\omega\tau_j)\, $ and
\BE{\vert  \hat{f}(\omega) \sum_{j=1}^\NN \exp(-i\omega\tau_j)  \vert^2 =
\vert \hat{f}(\omega) \vert^2 \left(  \NN + 2 \sum_{j=1}^\NN\sum_{k=1}^\NN\, \cos\omega(\tau_k - \tau_j) \right)\label{key}}

Hence for  a train of such equidistant pulses  with  spatial separations $cT_0>0$
  $$\dUU(\omega)=K F_0^2 \vert \hat{f}\vert^2(\omega)\,\,  \LL(\omega)$$
where $\LL(\omega)= \frac{1- \cos(\omega T_0 \NN)}{1- \cos(\omega T_0)}$ and $\dUU$ is independent
of $z$.

The function $\LL$ is bounded ($ 0 \le \LL(\omega)\le  \NN^2 $) with maxima at $\omega=\omega_j\equiv \frac{2\pi j}{T_0}$  but the  detailed behaviour of $\dUU(\omega)$ depends on the single pulse structure defined by $f$. Thus if $$ f(z-c(t-T_j))=\exp\left(-\left(\frac{z-c(t-T_j)}{\sigma_z}\right)^2 \right )$$ describes the structure of the $j-th$ pulse in the train with $T_j=j\,T_0$ one finds
  $$ \dUU(\omega)= KF_0^2\, \frac{\pi\sigma_z^2}{c^2} \ \WW(\omega)$$
where
  $$\WW(\omega)=\exp\left(-\left(\frac{\omega\sigma_z}{\sqrt{2} c}\right)^2\right)\, \frac{1-\cos(\omega T_0 \NN)}{1-\cos(\omega T_0)}$$

The modulation of $\LL$ by $\vert \hat f\vert^2$ means that the maxima of $\LL\,\vert \hat f\vert^2 $ are shifted from $\omega=\omega_j$. For typical charged bunches described by the above Gaussian form for $f$ the shift is unlikely to be experimentally detectable. However the points where $\omega=\omega_j$ determine the  first order contact points of the curve
$\dUU(\omega)$ with the envelope curve $$ \EE^{+}(\omega)\equiv K F_0^2\,\vert \hat f \vert^2(\omega) \NN^2$$
i.e. $$\underset{\omega\to\, \omega_j}{Lim}\,\, \left(  \frac {\EE^{+}(\omega)}  {\dUU(\omega)}\right) =1 $$

     $$\underset{\omega\to\, \omega_j}{Lim}\,\, \left(  \frac{(\EE^{+})^\prime (\omega)} {(\dUU)^\prime (\omega)}\right)=1 $$

The general features of the spectrum of $\dUU$ for the choice of  Gaussian $f_j$ are sketched in figure 1. It should be stressed that the electromagnetic fields generated by the non-stochastic source under consideration are fully \lq\lq coherent\rq\rq. The maxima in $\dUU$ are produced by constructive interference of the fields associated with the regular structure  in the train that is maintained during its ultra-relativistic motion. The spacing of adjacent  maxima of $\dUU(\omega)$  produced by bunches in an accelerator differ imperceptibly from the spacing $\frac{2\pi}{T_0}$ of adjacent points where $\dUU$ is tangent to $\EE^{+}$. Thus the longitudinal spatial separation  $c T_0$ of the maxima in this idealized bunch containing a structure with $\NN$  equidistant peaks is immediately visible in the electromagnetic energy spectrum.

\begin{figure}[ht]
\centering
    \subfigure{
    \includegraphics[scale=0.5]{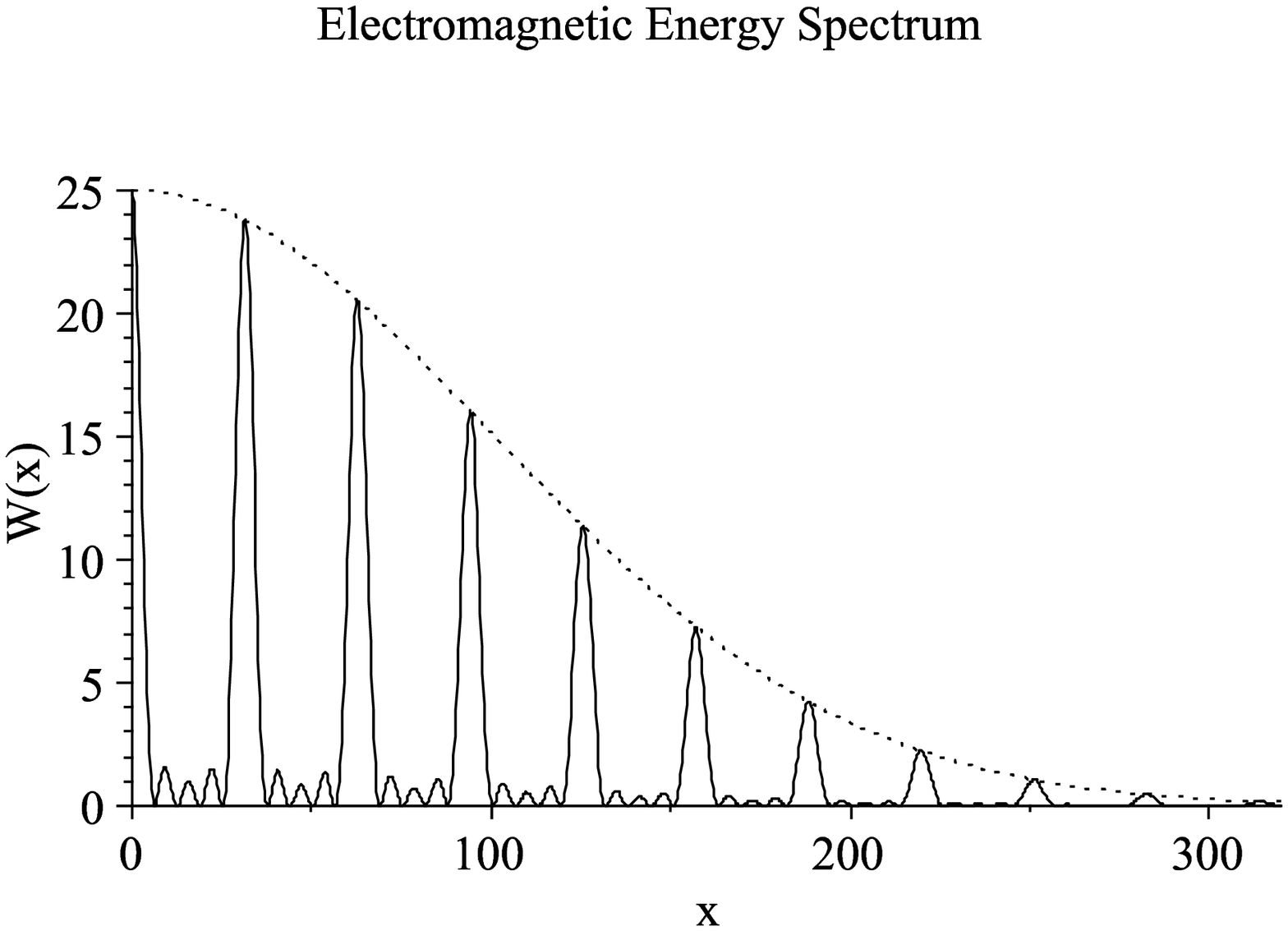}
    \label{fig:subfig1a}
    }
    \subfigure{
    \includegraphics[scale=0.3]{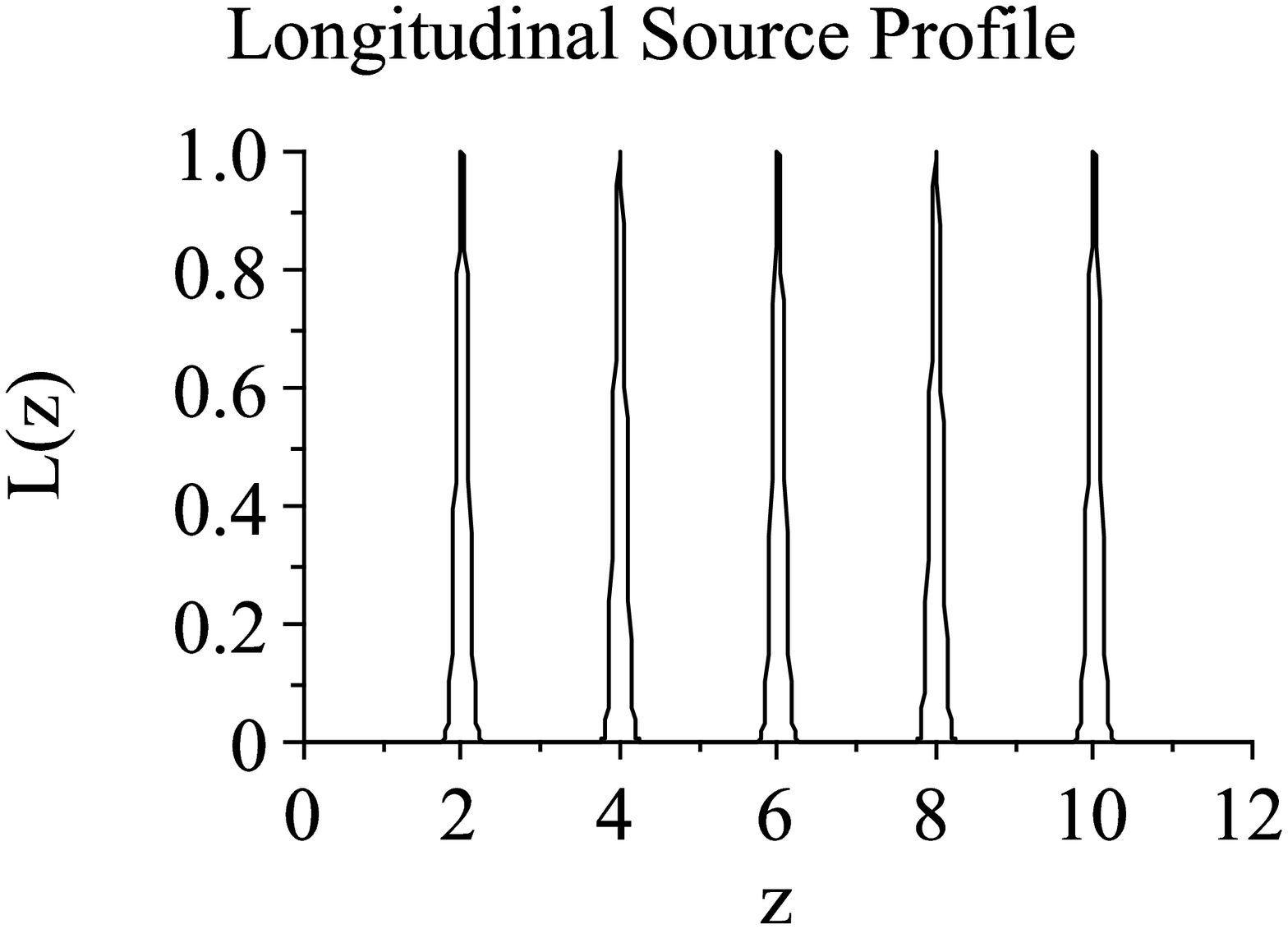}
    \label{fig:subfig1b}
    }
     \caption[figure3]{The rescaled
electromagnetic energy spectrum with $x=T_0 {\cal N} \omega$:\newline $ \it{W}(x)= \WW(\frac{x}{T_0\,{\mathcal N}}) $
associated with the longitudinal source profile \newline
 $L(z)\equiv \frac{F(z)}{F_0}=\sum_{j=1}^{\mathcal N}exp\left( -\left( \frac{z-c\,jT_{0}}{\sigma_z}%
\right) ^{2}\right). $
 For purposes of visualization
 $
 \sigma_{z}=\frac{1}{20%
}cT_{0},\,\,\mathcal{N}=5,\,\,T_{0}=1,\,\,c=1
 $.
   The upper dotted curve denotes the envelope $\frac{\EE^{+}(x)}{ (F_{0}^{2}K)}$.
 }
 \end{figure}

\section{{Stochastic Effects}}

To see the effects of randomization on these interference maxima  a simple stochastic model of a charged bunch containing $\NN$ identifiable random variables will be adopted. Instead of fixing the separation of the $\NN$  peaks in $\rho^{\parallel}$ they will be distributed according to some probability measure $P^\NN$. We choose as random variables $T_1,T_2,\ldots T_\NN$  and assume that $$P^\NN(T_1,T_2,\ldots T_\NN)=\Pi_{j=1}^\NN\, P(T_j)$$ where $P(T)$ is a probability distribution for a single random variable $T$. The expectation value of any function $\OO$ of $T_1,T_2,\ldots T_\NN$ will be denoted $\EEE_{P^\NN}(\OO)$ where
$$ \EEE_{P^\NN}(\OO) = \int_{R^\NN}  \OO(T_1,T_2,\ldots T_\NN) P^\NN( T_1,T_2,\ldots T_\NN) \,dT_1\,dT_2\ldots dT_\NN  $$

In particular it follows from (\ref{key}) that the expectation value
of the spectral energy distribution is
  $$ \EEE_{P^\NN} \left(\dUU(\omega)\right) = K F_0^2 \vert  \hat f \vert^2(\omega) \,\,  \EEE_{P^\NN} (\LL(\omega))  $$

where $$ \EEE_{P^\NN} (\LL(\omega)) = \NN + (\NN^2-\NN)\,\,\vert \int_{-\infty} ^\infty
P(T)\,e^{i\omega T}\,dT  \vert^2  $$ The magnitude of $(\NN^2-\NN)$ times  the modulus squared  of
the Fourier transform of  $P(T)$ in this expression  relative to $\NN$  determines the nature of
the expectation value of the spectral energy as a function of $\omega$. This expectation value is
now  bounded above {\it and below} by two distinct envelopes that vary with $\omega$.  Within
these envelopes one may classify local maxima as major and minor (see figure 2).  The width of the
first local dominant maxima in the $\omega$ spectrum of the expectation  value is directly related
to the overall scale of the spatial size of the bunch  source as determined by the probability
distribution $P(T)$   while the separation between  adjacent local major maxima in $\dUU$  is
determined by the structure of $P(T)$. These features  are illustrated  below where one notes that
the bounding envelopes have single maxima at $\omega=0$ in the ratio $\NN^2 : \NN$. For large
$\NN$ this is then close to the ratio of the first few ratios of ( local maxima : local minima) of
$\dUU$. Following tradition it is natural to refer to fields that contribute to the first few
major local maxima of $\dUU$ as exhibiting \lq\lq stochastic coherence\rq\rq.  The relation
between the non-zero frequency at which the first local minimum of $\dUU$ occurs  (or the
frequency beyond which $\dUU$ lies close to the lower bounding envelop) and the spatial
distribution of charge in $\rho^{\parallel}$ is a stochastic one depending on the structure of
$P(T)$. These general features are illustrated as follows.

\begin{itemize}

\item  If   $\NN$ identical pulses in $\rho^{\parallel}$, each with the above longitudinal Gaussian profile $f$,   are,  for some constant $\Gamma$, independently distributed according to $P^\NN$ with:
$$ P(T)= \frac{1}{\Gamma} \quad \mbox{for  }  -\frac{\Gamma}{2} \le T \le \frac{\Gamma}{2}$$
and zero elsewhere then
$$\EEE_{P^\NN}\left(\dUU(\omega)\right)= K F_0^2\, \vert \hat f \vert^2(\omega) \left( \NN + (\NN^2- \NN) \left(\frac{\sin(\omega\Gamma/2)}{   \omega\Gamma/2} \right)^2    \right).$$
The bounding envelopes (see figure 2) are the curves $\EE^{-}(\omega)=K F_0^2\, \NN\,\vert \hat f \vert^2(\omega)$
and $\EE^{+}(\omega)=K F_0^2\, \NN^2\,\vert \hat f \vert^2(\omega)$:

$$\EE^{-}(\omega)  \le  \EEE_{P^\NN}\left(\dUU(\omega)\right)  \le \EE^{+}(\omega)   $$

It is natural to designate  the Fourier components of fields as \lq\lq stochastically incoherent\rq\rq if they contribute to  the expectation $\EEE_{P^\NN}\left(\dUU(\omega)\right)$ in the vicinity of  the lower bounding envelope $\EE^{-}(\omega)$. This expectation value  first touches $\EE^{-}(\omega)$ at approximately $\omega= \frac{2\pi}{\Gamma}$ and $\Gamma$ determines the average spatial length of $\rho^{\parallel}$.

\begin{figure}[ht]
\centering
    \subfigure{
    \includegraphics[scale=0.5]{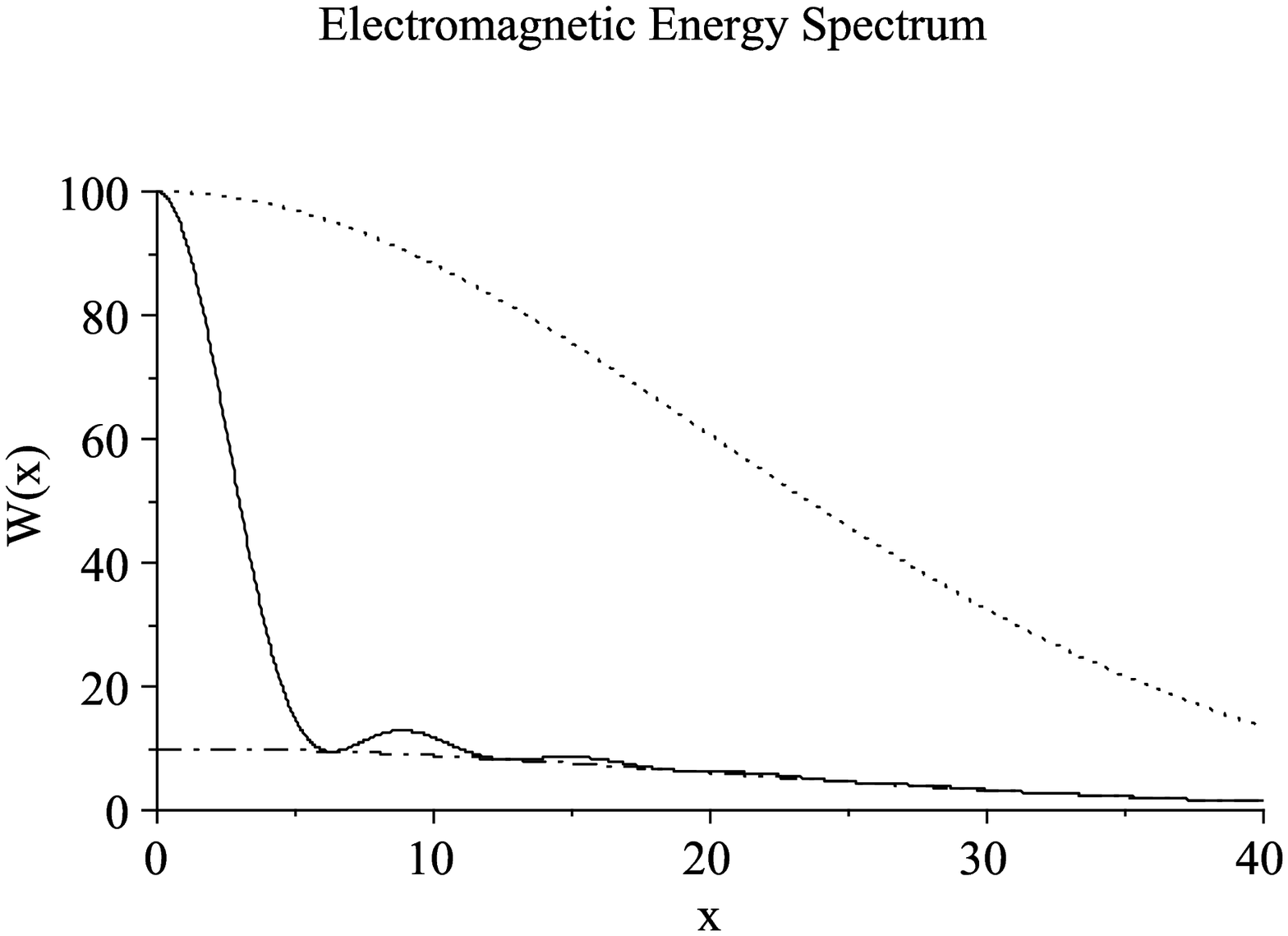}
    \label{fig:subfig2a}
    }
    \subfigure{
    \includegraphics[scale=0.3]{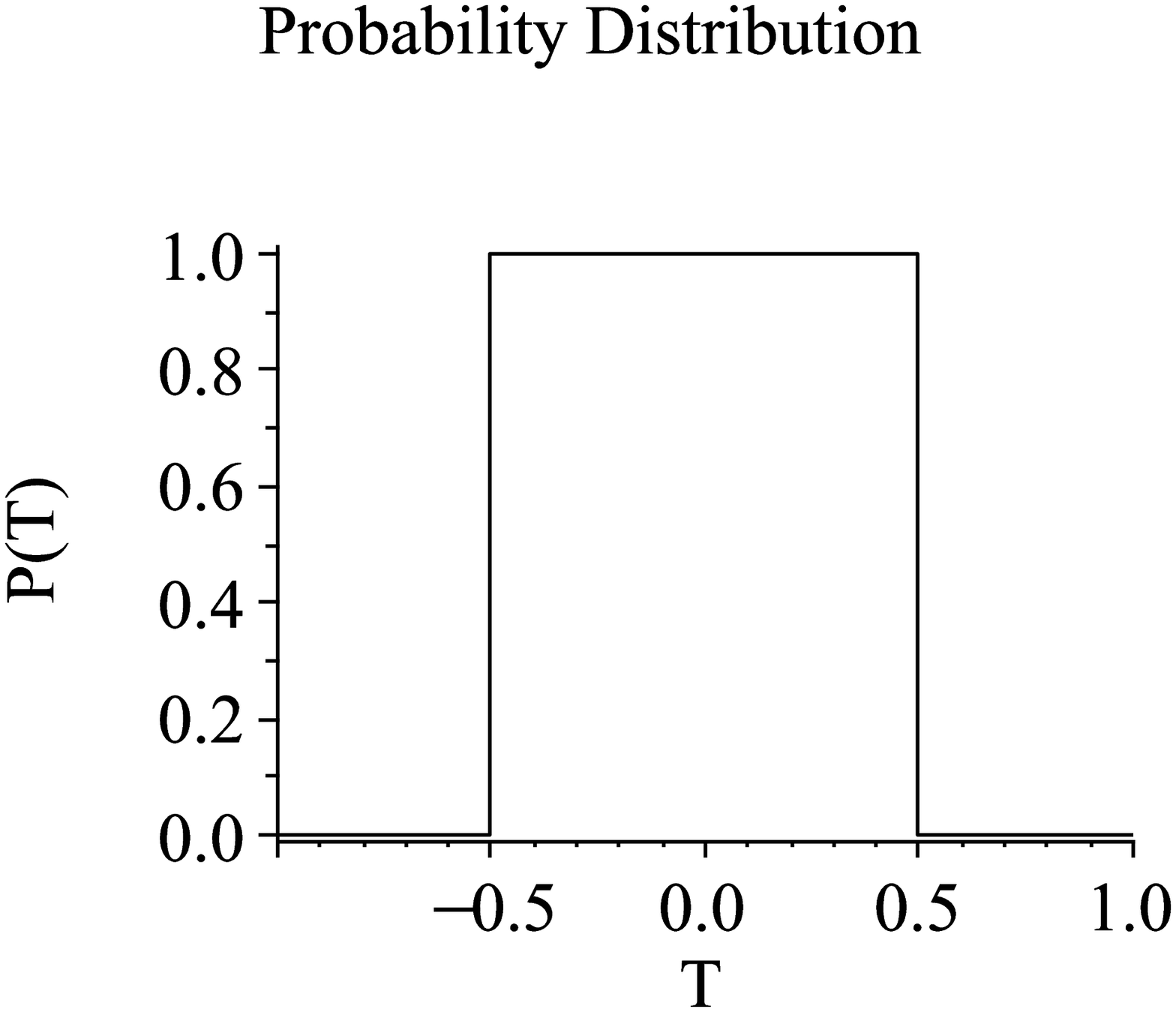}
    \label{fig:subfig2b}
    }
   \label{fig:figure2} \caption[figure2]{The rescaled
electromagnetic energy spectrum with $ x=\Gamma \omega $:\newline $ \it{W}(x)=
E_{P^\mathcal{N}}\left( \frac{d\mathcal{U}}{d\omega }\left(\frac{x}{\Gamma} \right)\right) /
(F_{0}^{2}K) $ associated with the longitudinal probability distribution \newline $P(T)=1/\Gamma
\mbox{ for }-\Gamma /2\leq T\leq \Gamma /2. $  For purposes of visualization
 $
 \sigma _{z}=\frac{1}{20}c\Gamma,\,\,\Gamma =1,\,\,c=1,\,\,\mathcal{N}=10
 $.
   The upper dotted curve denotes the envelope $\frac{\EE^{+}(x)}{ (F_{0}^{2}K)}$ and the lower one the envelope $\frac{\EE ^{-}(x)}{(F_{0}^{2}K)}$.
 }
 \end{figure}

\item  If, for some integer $n$ and constant $\kappa$,  $\NN$ identical pulses in $\rho^{\parallel}$, each with the above longitudinal Gaussian profile $f$,   are independently distributed according to $P^\NN$ with:
$$ P(T)= \frac{1}{(2n+1) \AA \sqrt{\pi}} \sum_{j=-n}^n \exp\left(-\left(\frac{T-j\tau}{\AA}\right)^2   \right)   \quad \mbox{for  }  -\infty \le T \le \infty$$ then
$$\EEE_{P^\NN}\left(\dUU(\omega)\right)= K F_0^2\, \vert \hat f \vert^2(\omega)\left(
\NN + (\NN^2-\NN)\frac{1}{(2n+1)^2}\exp\left(  -
\left(\frac{\omega\AA}{\sqrt{2}}\right)^2\right)
\frac{1-\cos(\omega\tau (2n+1))}{1-\cos(\omega \tau)} \right)$$

Such a distribution $P(T)$ offers a means to model the electromagnetic energy associated with a
bunch containing $\NN$ constituents with $2n+1 $ micro-bunches \cite{stupakov} distributed
stochastically among them according to equally spaced Gaussian distributions with parameters
$\kappa$ and $\tau$.

In this case the bounding envelopes (see figure 3) are the curves $$\EE^{-}(\omega)=K F_0^2\, \NN\,\vert \hat f \vert^2(\omega),$$
 $$\EE^{+}(\omega)=K F_0^2\, \vert \hat f \vert^2(\omega) \left(\NN -(\NN^2-\NN) \exp(-\frac{\omega^2\AA^2}{2}  )    \right) \,$$
 and
$$\EE^{-}(\omega)  \le  \EEE_{P^\NN}\left(\dUU(\omega)\right)  \le \EE^{+}(\omega)   $$
For $\omega \neq 0$ the envelope $\EE^{+}(\omega)$ here lies below the envelope  $\EE^{+}(\omega)$ for the previous distribution in figure 2. Hence the overall effect of the micro-structure introduced in this $P(T)$ is to suppress  the magnitude of the $j>0$  local major maxima  near $\omega= \frac{2\pi j}{\tau}$ in the expectation values of $\dUU$.

\begin{figure}[ht]
\centering
    \subfigure{
    \includegraphics[scale=0.5]{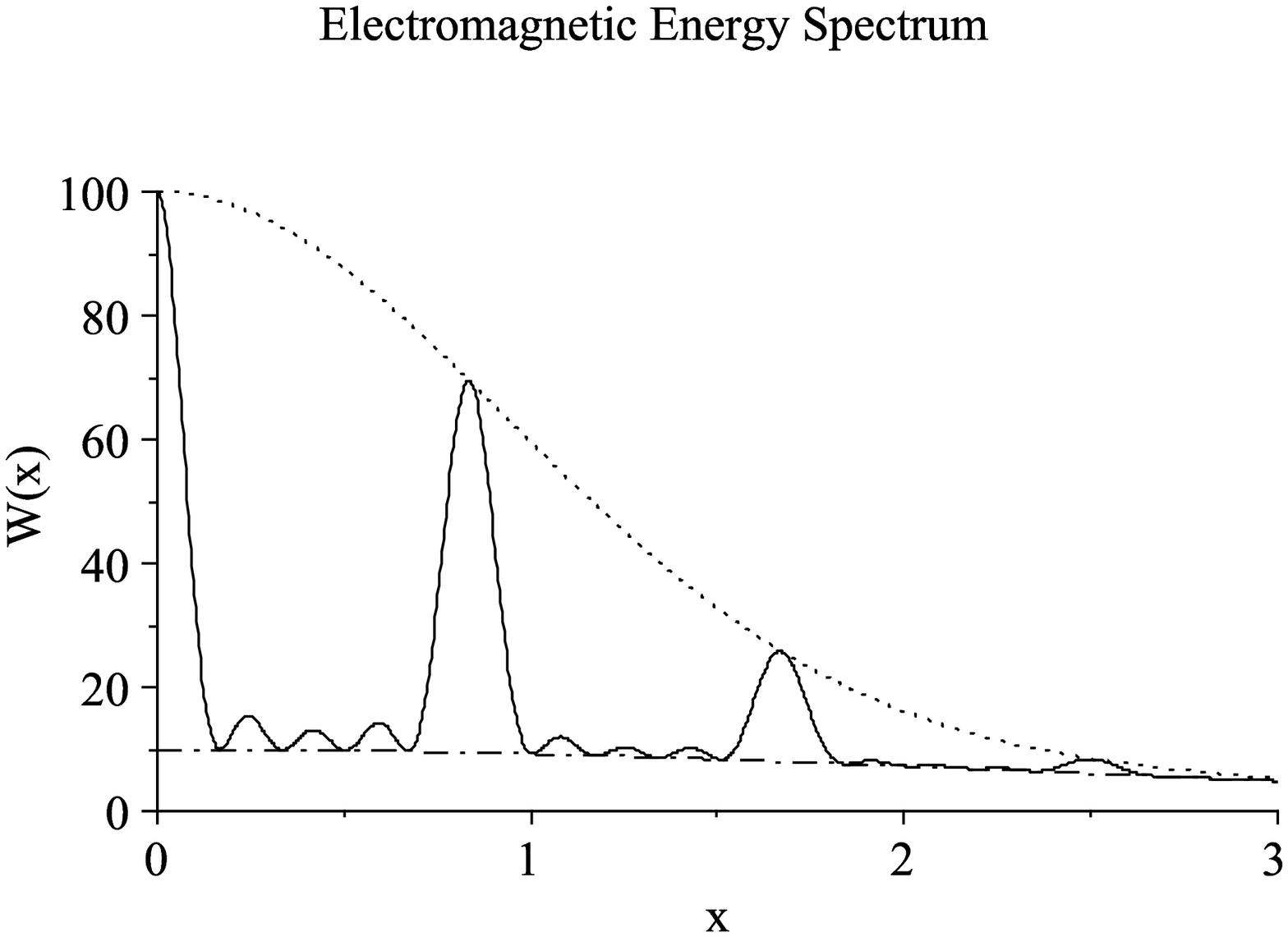}
    \label{fig:subfig3a}
    }
    \subfigure{
    \includegraphics[scale=0.3]{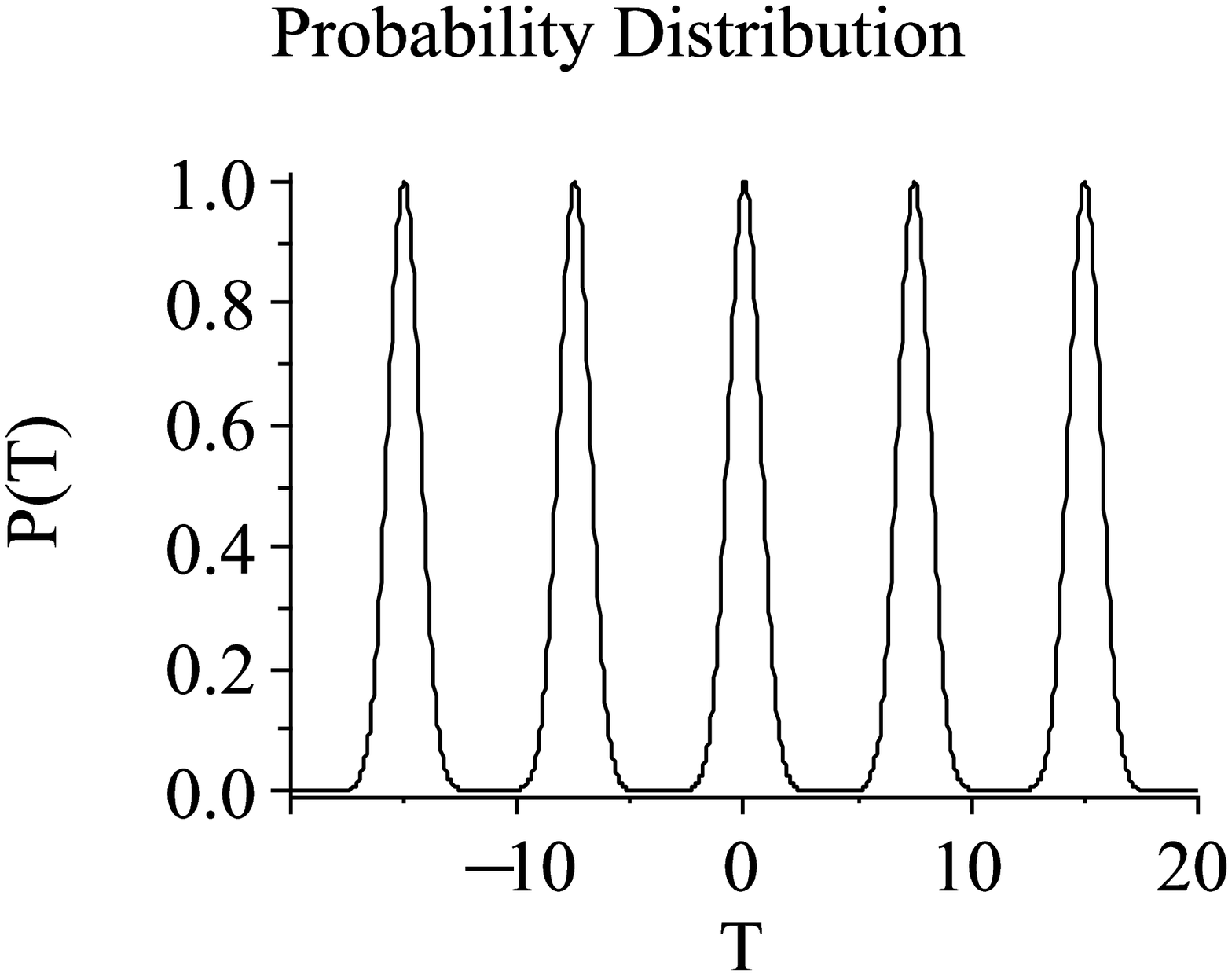}
    \label{fig:subfig3b}
    }
\label{fig:figure3} \caption[figure3]{The rescaled electromagnetic energy spectrum with $x=\kappa \omega  $: $ \it{W}(x)=
E_{P^\mathcal{N}}\left( \frac{d\mathcal{U}}{d\omega }\left(\frac{x}{\kappa} \right)\right) /
(F_{0}^{2}K) $ associated with the probability distribution \newline $P(T)=\left( (2n+1)\kappa
\sqrt{\pi }\right) ^{-1}\sum_{j=-n}^{n}\exp \left( -(T-j\tau )^{2}\kappa ^{-2}\right). $ \newline
For purposes of visualization
 $\sigma _{z}=\frac{2}{5}c\kappa
,\,\,\tau =30\kappa (n-1)^{-1}\,\,\,,\,\mathcal{N}=10,\,\,n=2,\,\,\kappa
=1,\,\,c=1$.
  The upper dotted curve denotes the envelope $\frac{\EE^{+}(x)}{ (F_{0}^{2}K)}$ and the lower one the envelope $\frac{\EE ^{-}(x)}{(F_{0}^{2}K)}$.
   }

\end{figure}

\end{itemize}

\section{{Conclusions}}

A general discussion  of the electromagnetic energy spectra generated by ultra-relativistic
distributions of charge in a perfectly conducting straight cylindrical beam pipe has been given. A
detailed model has been constructed for particular distributions  where the electromagnetic fields
have been calculated ab-initio by solving analytically the initial-boundary value problem for
Maxwell's equations in the beam pipe. The effects of the pipe geometry and the source proper
charge distribution on the field structure have been explored for different stochastic models  and
the resemblance of this structure to CSR behaviour in a cyclic machine pointed out.  Further
aspects of such dynamically enhanced field effects can be found in \cite{AlisonPhD}.  Our results
offer a viable experimental means for inferring properties of the longitudinal charge distribution
of bunches with micro-structure in  ultra-relativistic motion in straight segments of a beam pipe
from observation of the associated electromagnetic energy spectra. They are also of relevance to
design criteria where the coherence effects of such fields play a significant role.

\subsection*{Acknowledgments}
The authors are grateful to D. A. Burton, A. Cairns, S. P. Jamison
and A. Wolski for valuable discussions and to the Cockcroft
Institute, STFC and EPSRC for financial support for this research.

\pagebreak[4]



\begin{thebibliography}{99}
\bibitem{schwinger_1949} J Schwinger,
{\it On the Classical Radiation of Accelerated Electrons}, Phys. Rev. {\bf 75} 1912, (1949).
\bibitem{schwinger_1946} J Schwinger,
{\it Electron Radiation in High Energy Accelerators}, Phys. Rev. {\bf 70} 798, (1946).
\bibitem{hofmann} A Hofmann,
{\it The Physics of Synchrotron Radiation}, (Cambridge University Press, 2004).
\bibitem{saxon_1954} J S  Nodvick, D S Saxon,
{\it Suppression of Coherent Radiation by Electrons in a Synchrotron}, Phys. Rev. {\bf 96} 180,
(1954).
\bibitem{saldin} E L Saldin, E A Schneidmiller, M V Yurkov
{\it Coherent Radiation of an Electron Bunch Moving in an Arc of a Circle}, Nuclear Instruments
and Methods in Physics Research A {\bf 398} 373-394, (1997).
\bibitem{jamison} S P Jamison, et al,
{\it Femtosecond Resolution Electron Bunch Profile Measurements}, Proceedings of EPAC 2006
Edinburgh Scotland 915-919, (2006).
\bibitem{jones} D S Jones,
{\it The Theory of Electromagnetism}, (Pergamon Press, 1964).
\bibitem{tucker1} R W Tucker,
{\it On the Effects of Geometry on Guided Electromagnetic Waves}, Theoret. Appl. Mech. {\bf 34}
1-49, (2007).
\bibitem{tucker2} S Goto, R W Tucker,
{\it Electromagnetic Fields Produced by Moving Sources in a Curved Beam Pipe}, J.Math Phys. {\bf
50} 063510, (2009).
\bibitem{pinsky}  M A Pinsky,
{\it Partial Differential Equations and Boundary-Value Problems with Applications}, (McGraw-Hill,
Inc.,  1991).
\bibitem{stupakov} G Stupakov, S Heifets,
{\it Beam Instability and Microbunching Due to Coherent Synchrotron Radiation}, Phys. Rev. ST
Accel. Beams {\bf 5} 054402, (2002).
\bibitem{AlisonPhD} A C Hale,
{\it Aspects of Dynamically Enhanced Electromagnetic Fields from Charged Relativistic Sources in a
Beam Pipe}, (PhD Thesis, Lancaster University, 2008).

\end{thebibliography}
\end{document}